\documentclass[aps,prl,groupedaddress,twocolumn,10pt]{revtex4-1}

\usepackage{graphicx, bm, amsfonts, amssymb, amsmath, amsthm, color}

\newcommand{\WF}{\mathbb{WF}}

\newcommand{\bea}{\begin{eqnarray}}
\newcommand{\eea}{\end{eqnarray}}
\newcommand{\nn}{\nonumber}
\newcommand{\be}{\begin{equation}}
\newcommand{\ee}{\end{equation}}

\theoremstyle{definition}
\newtheorem{defi}{Definition}
\newtheorem{pro}[defi]{Property}
\theoremstyle{plain}
\newtheorem{con}[defi]{Conjecture}

\newcommand{\Z}{\mathcal{Z}}
\newcommand{\CC}{\mathcal{C}}
\newcommand{\GG}{\mathcal{G}}

\newcommand{\R}{\mathbb{R}}
\newcommand{\C}{\mathbb{C}}

\newcommand{\calD}{{\mathcal D}}

\newcommand{\id}{\mathbf{1}}

\newcommand{\ZZ}{\mathcal{Z}}

\newcommand{\act}{\triangleright}

\newcommand{\hodge}{\ast}

\newcommand{\lalg}[1]{\mathfrak{#1}}  
\newcommand{\SU}{\mathrm{SU}}
\newcommand{\SO}{\mathrm{SO}}

\newcommand{\OO}{\mathrm{O}}

\newcommand{\Spin}{\mathrm{Spin}}

\newcommand{\su}{\lalg{su}}

\newcommand{\spin}{\lalg{spin}}
\newcommand{\dd}{\mathrm{d}}

\def\ra{\rangle}

\begin{document}

\title{Geometric asymptotics for spin foam lattice gauge gravity on arbitrary triangulations}
\author{Frank Hellmann}
 \email[Electronic Address: ]{Frank.Hellmann@aei.mpg.de}
 \affiliation{MPI for Gravitational Physics, Golm, Germany}

\author{Wojciech Kami{\'n}ski}
 \email[Electronic Address: ]{Wojciech.Kaminski@fuw.edu.pl}
 \affiliation{MPI for Gravitational Physics, Golm, Germany}

\date{\today}


\begin{abstract}
We study the behavior of holonomy spin foam partition functions, a form of lattice gauge gravity, on generic 4d-triangulations using micro local analysis. To do so we adapt tools from the renormalization theory of quantum field theory on curved space times. This allows us, for the first time, to study the partition function without taking any limits on the interior of the triangulation.

We establish that for many of the most widely used models the geometricity constraints, which reduce the gauge theory to a geometric one, introduce strong accidental curvature constraints. These limit the curvature around each triangle of the triangulation to a finite set of values. We demonstrate how to modify the partition function to avoid this problem. Finally the new methods introduced provide a starting point for studying the regularization ambiguities and renormalization of the partition function.
\end{abstract}

\maketitle


\section{Introduction}

Spin foam models for quantum gravity are a form of lattice gauge gravity. They are constructed by a modification of topological lattice theories \cite{ponzanoregge,Fukuma:1993hy,Baez:1997zt,Barrett:2008wh}. These are heuristic lattice quantizations of so called $BF$ theory \cite{Horowitz:1989ng}, whose only equation of motion is to enforce flatness of the connection. The intuition is to use so called simplicity constraints to restrict the Lagrange multipliers enforcing flatness. The restricted multipliers only enforce Ricci flatness, rather than full flatness, and we obtain general relativity \cite{Capovilla:1989ac,Capovilla:1991qb,Reisenberger:1996ib,Reisenberger:1997sk,Reisenberger:1998fk,Reisenberger:1998bn}.

This is done at the level of the discrete partition function on a fixed lattice \cite{Barrett:1997gw,Barrett:1999qw,Livine:2007ya,Livine:2007vk,Engle:2007uq,Engle:2007qf,Engle:2007wy,Freidel:2007py,Pereira:2007nh,Baratin:2010wi,Baratin:2011hp}, thus one only obtains discrete gravity this way. A fully satisfactory restriction to the geometric sector is not known \cite{Dittrich:2010ey,Engle:2011ps,Engle:2011un,Engle:2012yg}, however it was shown that the weights of the partition function are approximated for large quantum numbers by the Regge action of discrete gravity\cite{Barrett:1993db,Barrett:1998gs,Barrett:2002ur,Baez:2002rx,Conrady:2008ea,Conrady:2008mk,Barrett:2009gg,Barrett:2009mw,Barrett:2009as,Barrett:2010ex,Hellmann:2010nf,Hellmann:2011jn,Conrady:2009px}, raising the hope that a continuum limit towards full gravity might be feasible.

Up until now no method for analyzing the full dynamics of the theory that did not rely on the large quantum number approximation throughout the partition function was available. Furthermore the partition functions obtained in this way all require regularization, and the ambiguities of these regularizations remained ill understood. This is a crucial question as, if the ambiguities proliferate as the lattice gets finer, they would render any continuum limit completely unpredictive.

In this letter we show that the notion of the wave front set of a distribution, introduced in the context of the renormalization analysis of quantum field theory in curved space time, can be adapted towards the study of both of the above issues.
 
%
%

\section{The partition function}\label{sec:PartFunc}

The partition function for spin foam models we will use is introduced and studied in \cite{Bahr:2012qj,Dittrich:2012er}. It is defined on a triangulation of a manifold. Conventionally we label the elements of the triangulation with the corresponding elements in the dual of the triangulation. That is, a vertex for every 4-simplex, an edge for every tetrahedron, and so on. We will only require the 2-skeleton of the dual, putting us in the same setting as lattice gauge theory, with vertices, edges and faces (plaquettes).

Let $\CC$ be the 2-complex given by the 2-skeleton of the dual of a triangulation, with vertices $v$, edges $e$ and faces $f$, and fiducial orientation and base vertex on each face. Call the graph of edges and vertices in the boundary of the 2-complex $\Gamma$. We have one $\Spin(4)$ group element $g_{ev} = g_{ve}^{-1}$ for every half edge, labeled by an adjacent pair of edge $e$ and vertex $v\in e$. These are interpreted as the holonomies from the middle of the edge to the vertex, defining the discrete connection. We further have one group element for every adjacent pair of face $f$ and edge $e \in f$, called $g_{ef}$. From these we construct two new group elements by going around the face:

\bea
g_f &=& g_{ve}g_{ef}g_{ev'}g_{v'e'} \dots g_{e^{(n)}f}g_{e^{(n)}v}\nn\\
\tilde{g}_f &=& g_{ve}g_{ev'}g_{v'e'} \dots g_{e^{(n)}v}\nn\\
\eea

Here $v$ is the fiducial base vertex and the order of group elements is taken with respect to the fiducial orientation of the face. We can then define the partition function:

Let $E(g)$ be a distribution on $\Spin(4)$ satisfying $E(g)=E(hgh^{-1})$, $\forall h \in \SU(2)_{\mbox{diag}} \subset \Spin(4)$. Let $\omega(g)$ be a distribution obtained by acting with a pseudo-differential operator on $\delta(g)$ satisfying $\omega(g)=\omega(g'g{g'}^{-1})$, $\forall g' \in \Spin(4)$. Then the spin foam partition function on $\CC$ is given by

\be\ZZ(\CC) = \int_{e \notin \Gamma} \dd g_{ev} \int \dd g_{ef} \prod_{f} \omega(g_{f}) \prod_{e \subset f} E(g_{ef}) \label{eq:PartFunc} \ee

The partition function thus is a distribution in the universal boundary Hilbert space associated to $\Gamma$, which was introduced in \cite{Bahr:2012qj,Dittrich:2012er}. In the companion paper \cite{HellKami} we will give the analysis for the more familiar projected spin network space \cite{Alexandrov:2002xc,Livine:2002ak,Dupuis:2010jn}.

Different choices of $E$ and $\omega$ lead to different models. In particular $E=\omega=\delta$ is the special case of $BF$ theory. The $g_{ef}$ are eliminated from the partition function and the $g_{ev}$ are forced to be the holonomies of a flat connection by $\omega=\delta$.

The role of the $E$ function is to allow the $g_{ef}$ to differ from $\id$ in such a way that some curvature is allowed in the discrete connection $g_{ev}$. Thus the $E$ encode the simplicity constraints.

\section{Wave front sets}

In order to extract the geometric content of the partition function \eqref{eq:PartFunc} we introduce a new tool into the study of lattice gauge gravity, the wave front set of a distribution \cite{lars,grigis1994microlocal}. The wave front set is a subspace of the cotangent bundle over the space on which the distribution is defined. Interpreting the distribution as a (generalized) wave function, it can be understood intuitively as the subspace of phase space on which the distribution is peaked in the limit of larger momenta.

We will now give the precise definition. Let $M$ be a smooth compact manifold and ${\calD}(M)$ the distributions over $M$. We denote $\{0\}$ the zero section of the cotangent bundle $T^*M$.
%
The wave front set $\WF(A) \subset T^*M$ of $A \in \calD(M)$ is defined as the complement of the set of
elements $\{(x,p)\in T^*M\setminus\{0\}\}$ such that there exists a local coordinate
patch $U\times V$ containing $(x,p)$ with
\begin{align}
\forall \phi\in C^\infty_0(U),& \phi(x) \neq 0, \; \tilde p \in V\colon\nn\\ &\int_U e^{i\lambda \tilde p\tilde{x}} \phi(\tilde{x})A(\tilde{x})\dd\tilde{x}=O(\lambda^{-\infty})
\end{align}

We always have $\{0\}\subset \WF(A)$, this is a minor deviation from the conventions in \cite{lars,grigis1994microlocal} which always exclude $\{0\}$. This change of convention significantly simplifies our bookkeeping. $\WF(A)$ is a geometric cone in $T^*M$.

\subsection{The construction properties}

Wave front sets behave well under composition in two crucial ways that we will explain now. The first enables us to obtain parallel transport equations:

\begin{pro}[Parallel transport]\label{pro-parallel} Let $A\in \calD(G)$ where $G$ is a Lie group. Define $\tilde{A}\in\calD(G\times\cdots \times G)$
\begin{equation}
 \tilde{A}(g_1,\ldots, g_n)=A(g_1\cdots g_n)
\end{equation}
then we can show that
\begin{align}
 \WF(\tilde{A})=\{& (g_1,p_1,\ldots,g_n,p_n)\colon
 (g_1\cdots g_n,p_n)\in \WF(A),\nn\\& \forall_{i < n} \;
p_i= g_{i+1} \act p_{i+1}\}
\end{align}
\end{pro}

This property allows us to obtain the wavefront set of $\omega(g_f)$ as a distribution on the $g_{ef}$ and $g_{ev}$.

The second property is slightly more complicated. Consider distributions $A_i$ on spaces $M_i \times M$. We will be interested in understanding the convolution product when integrating out $M$. Thus we call the elements $(x, p) \in T^* M$ the interior, and the elements of $(x_i, p_i) \in T^* M_i$ the boundary variables. We use $\partial$ for the restriction to the boundary variables. We denote as the interior closed subspace $W_{\mbox{icl}}$ of a space $W \subset \times_i (T^*(M_i \times M))$ the elements in $W$ such that the interior position $x$ in each factor agrees and that the interior momenta sum to zero. We call the momentum in the $i$th factor $p^i$, and the boundary part of it $\partial p^i$.
%

\begin{pro}[Closure]\label{pro-closure} If \be\{(\times_i \WF(A_i))_{\mbox{icl}}| \partial p^i = 0\} = \{(\times_i \WF(A_i))_{\mbox{icl}}| p^i = 0\},\ee then $A=\int dm A_1\cdots A_n$ exists as a distribution and
\begin{equation}
\WF(A)\subset \partial (\times_i \WF(A_i))_{\mbox{icl}}
\end{equation}
\end{pro}

A set of elementary properties from which these two can be derived is contained in the supplemental material.

\section{The wave front set of the partition function}

In order to obtain the wave front set of \eqref{eq:PartFunc} we will first need the wave front sets of $\omega$ and $E$. First note that any delta function $\delta_N$ that peaks on a submanifold $N \subset M$, has as its wave front set the space of points $(x,p)$ in $T^*M$ for which $x \in N$ and $p$ annihilates the tangent vectors to $N$. If we operate with a pseudo differential operator on a distribution the new wave front set is a subset of the old one. Thus we immediately have that
\be
\WF(\omega) \subset \{(\id,p), \forall p\} \cup \{0\}.
\ee

For the $E$ function we specialize to the function $E^\gamma_{\mathrm{EPRL}}$ of the model of Engle, Pereira, Rovelli and Livine \cite{Engle:2007wy}, which depends on a rational parameter $\gamma \neq \pm 1$. We can identify a set of differential equations that are solved by this function, and then analyze the principal symbol of the corresponding differential operator \cite{HellKami}. Using this we obtain that \begin{align}
&\WF(E^\gamma_{\mathrm{EPRL}})=\nn\\&\{(g,p)\colon N^0 \cdot (1 - \gamma \hodge)p = 0, g = \exp{\xi \hodge \hat{p}}, \xi\in {\mathbb R}\},
\end{align} where $\hodge$ is the Hodge dual on $\spin(4)$, $N^0 = (1,0,0,0)$ and $\hat p$ is the normalized Lie algebra element.

We further conjecture that the wave front set of the model of Freidel and Krasnov \cite{Freidel:2007py} is the same: $\WF(E^\gamma_{\mathrm{EPRL}}) = \WF(E^\gamma_{\mathrm{FK}})$. The wave front set of this model without $\gamma$ is obtained by setting $\gamma = 0$ in the above formula. Further note that for the original model of Barrett and Crane \cite{Barrett:1997gw} we have $E_{\mathrm BC} = \delta_{\SU(2)_{\mathrm diag}}$ and thus,
\be \WF(E_{\mathrm{BC}})=\{(g,p)\colon N^0 \cdot * p = 0, g \in \SU(2)_{\mathrm diag} \}.
\ee
The model of Baratin and Oriti \cite{Baratin:2011hp} can be seen as a version of this that incorporates $\gamma$.

\emph{The wave front set of $\ZZ$.} We can now state the wave front set of the partition function $\ZZ^\gamma$ with $E^\gamma_{EPRL}$. We will give the result in a form that anticipates the geometric interpretation as much as possible. Take $g_{ev}$, $g_{ef}$ as above and introduce $\spin(4)$ elements $p^v_{ee'}$ and $p^e_{vf}$ for $e$,$v$,$e'$,$f$ all adjacent to each other. Then we call $\GG^\gamma_\CC$ the solution space of the equations 
\bea\label{eq:G-VertexEquations}
p^v_{ee'} &=& - p^v_{e'e},\nn\\
p^v_{ee'} &=& g_{ve} \act p^e_{vf}.
\eea
at the vertex,
\bea\label{eq:G-twisted-simplicity}
N^0 \cdot (1-\gamma \hodge)p^e_{vf} &=& 0,\nn\\
\displaystyle \sum_{f \ni e} p^e_{vf} &=& 0,\nn\\
p^e_{vf} &=& - p^e_{v'f},
\eea
and
\be\label{eq:G-EdgeEquations}
\exists \xi_{ef} \; \mbox{s.t.} \; g_{ef} = \exp(\xi_{ef} \hodge \hat{p}^e_{vf}),
\ee
at the edge, and
\be\label{eq:G-FaceEquations}
g_f = \id\; \mbox{or}\; p^e_{vf} = 0,
\ee
on the face.

Recall that $\Gamma \subset \CC$ is the boundary graph. The boundary vertices, dual to tetrahedra, have a unique interior edge which we call $e(v)$. Thus every $g_{ev}$ at a boundary edge $e \in \Gamma$ has a unique associated $\spin(4)$ element $p^v_{ee(v)}$. Call $\partial_{\Gamma}$ the projection of $\GG_{\CC}$ on the subspace $(g_{ev},p^v_{ee(v)}) \in \prod_{ev, e \in \Gamma} T^*M$.

\emph{Main Result:} We find that if $p = 0\; \forall p \in \partial_{\Gamma}\GG^\gamma_\CC$ implies $p=0 \; \forall p \in \GG^\gamma_\CC$, then $\ZZ^\gamma(\CC)$ exists as a distribution and \be\WF(\ZZ^\gamma) \subset \partial_{\Gamma}\GG^\gamma_\CC\;.\label{prop-WF(Z)}\ee

The wave front set of the integrand and the partition function itself are subsets of $(\times_{ev} T^*\Spin(4))\times(\times_{ef} T^*\Spin(4))$. That is, we have group elements $g_{ev}$ and $g_{ef}$ and Lie algebra elements $p_{ev}$ and $p_{ef}$. As noted before we can identify $p_{ev} = p^v_{ee(v)}$ on the boundary. Correctly identifying $p_{ev}$ and $p_{ef}$ with the $p^v_{ee'}$ and $p^e_{fv}$ in the interior is more subtle and will be discussed in detail in \cite{HellKami}. Next to applying the constructive properties \ref{pro-closure} and \ref{pro-parallel} this is the main part of the prove of proposition \ref{prop-WF(Z)}.

\section{Geometric interpretation}

We now need to give a geometric interpretation of the solution space $\GG_\CC^\gamma$. We begin with $\GG_\CC^0$. By the reconstruction theorems of \cite{Barrett:1997gw,Barrett:2009gg}, we obtain, up to non-degeneracy assumptions and symmetries, a geometric 4-simplex per vertex which we call $\sigma^v$ with boundary tetrahedra $\tau^v_e$ from equations \eqref{eq:G-VertexEquations} and \eqref{eq:G-twisted-simplicity}.

Note that the $g_{ev}$ furnish the discrete connection in the sense that $g_{ev} \tau^v_e = \tau^e_v$. The faces of $\tau^e_v$ are encoded by the $p^e_{vf}$. In the parallel transport equations that occur in the wave front set of the $\omega$ we see that $g_{ef}$ acts on these faces. For the model of Barrett and Crane $g_{ef}$ can change these $p$ and the tetrahedral geometry can change along the edge. This way we recover the known ultralocality problem of this model. In the case of $\ZZ^\gamma$ on the other hand $g_{ef}$ stabilize the $p^e_{vf}$, $\tau^e_v = \tau^e_{v'}$ and we obtain a continuous geometry throughout the manifold.

As we will discuss in more detail in the next section, the condition $g_{f} = \id$ fixes the sum of the $\xi_{ef}$ around a face to be proportional to the deficit angle $\Theta_f$ encoded in the product $\tilde g_f$ of $g_{ev}$ we gave before. A precise discussion of the geometry is given in the supplemental material. We see that the equations of $\GG^0_\CC$ reduce our variables to geometric configuration, up to the well studied ambiguities.

%

\subsection{Accidental curvature constraints.} Introducing the parameter $\gamma$ changes the picture in the geometric sector dramatically. Almost all geometric configurations in $\GG^0_\CC$ do not occur in $\GG^\gamma_\CC$.

To see this, note that given a solution from $\GG^\gamma_\CC$ we can apply the the (invertible) operator $T^\gamma = \frac1{\sqrt{1 + \gamma^2}} (1-\gamma \hodge)$ to the $p$, $p' = T^\gamma p$. As $T^\gamma$ is linear and commutes with the adjoint action of $\Spin(4)$, the $p'$ satisfy equations \eqref{eq:G-VertexEquations} and \eqref{eq:G-twisted-simplicity} for $\gamma = 0$. Thus we can again reconstruct the geometric simplices at each vertex, and $g_{ev}$ are the geometric connection in the same sense as in $\GG_\CC^0$.

However, the equation $g_f = \id$ now starts playing a dramatically different role. Note that $g_{ef}$ is given in terms of $p'^e_{vf}$ by $g_{ef} = \exp(\xi'_{ef} (\hodge + \gamma) \hat{p}'^e_{vf})$ where $\hat{p}'$ is again normalized and $\xi' = \frac{\sqrt {1+\gamma^2}}{1-\gamma^2} \xi$ is rescaled. It is straightforward to see that $g_{f}$ stabilizes $p'_f = \hat{p}'^v_{ee'}$ at the fiducial vertex $v$.

Thus it can be written as $g_f = \exp(\theta_1 p'_f + \theta_2 \hodge p'_f)$ and $g_f = \id$ decomposes into $\theta_1 = \theta_2 = 0$. Now in the geometric sector the $g_{ev}$ contribute the angle of the face holonomy, which is the negative deficit angle, $-\Theta_f$ to $\theta_2$, while the $g_{ef}$ contribute $\sum \xi'_{ef}$ to $\theta_2$ and $\gamma \sum \xi'_{ef}$ to $\theta_1$
\begin{equation}
\theta_2 = - \Theta_f + \sum \xi'_{ef}, \; \theta_1 = \gamma \sum \xi'_{ef}.
\end{equation}

For $\gamma = 0$ the equation on $\theta_1$ is redundant, and the condition on $\theta_2$ fixes the geometric meaning of $\sum \xi_{ef}$, as claimed before. However for $\gamma \neq 0$ we obtain the accidental curvature constraints:
\begin{equation}
- \theta_2 = \gamma \Theta_f = 0 \mod 4\pi.
\end{equation}

Thus only finitely many curvature values are allowed, unless we take $\gamma$ to $0$ first \cite{Magliaro:2011dz}.

One way to fix this issue is to replace $\omega(g_f)$ with a distribution $D(g_f,\tilde{g}_f)$, which depends also on the holonomy around the face $\tilde{g}_f = g_{ve} g_{ev'} \dots g_{e''v}$ directly, with the wave front set \begin{align} &\WF(D) = \{(g,\tilde{g}, p, \tilde{p})\colon\nn\\& p = \tilde{p}, \exists \xi\; s.t.\; g = \exp(\xi \gamma \hodge p), \tilde{g} = \exp(\xi p)\}.\end{align}

We conjecture that this can be achieved by a distribution defined in terms of $\SU(2)$ coherent states $n|n\ra = \frac12 |n\ra \in \C^2$, $n \in \su(2)$, $|n|=1$ as
\begin{align} D(g,\tilde{g}) = \sum_j \int& \dd n \,\dd m\, \langle n|g^+|n \rangle^{2j} \langle m|g^-|m \rangle^{2j}\times\nn\\&\times  \langle n|\tilde{g}^+|n \rangle^{2\gamma j} \langle -m|\tilde{g}^-|-m \rangle^{2 \gamma j}. \end{align}

\section{Regularization of the distribution}

The conditional of proposition \ref{prop-WF(Z)} that $p = 0$ for all $p \in \partial_{\Gamma}\GG^\gamma_\CC$ implies $p=0$ for all $p \in \GG^\gamma_\CC$, will not be true for most $\CC$. Then the partition function diverges and requires regularization. This is already a subtle problem for BF theory \cite{Bonzom:2010ar,Bonzom:2010zh,Bonzom:2011br,Bonzom:2012mb,Bahr:toappear}. The simplicity constraints should reduce the number of divergences we are dealing with though.

On a technical level the problem we face is to extend the distributions in question to the configurations on which they are ill defined. The issue is then what ambiguities arise in this extension. Instead of using property \ref{pro-closure} directly we can consider the extension problem of the integrand of \eqref{eq:PartFunc}. As extending a distribution does not usually change the location of it's wave front set it appears reasonable that the extension of the integrand still will include the condition $g_f = \id$, as well as restricting $g_{ef}$ to be generated by Lie algebra elements of the form $\hodge p + \gamma p$. As argued in the preceding section, together with the geometricity of the connection these imply the accidental curvature constraints. Thus we make the following conjecture:

\begin{con}[Regularized flatness]
Any regularization of $\ZZ(\CC)$ for which the $g_{ev}$ retain their interpretation as a discrete geometric connection contain the curvature constraint $\gamma \Theta_f = 0 \mod 4\pi$ for the bulk connection.
\end{con}

\section{Discussion}

In this letter we demonstrated a powerful new method for the analysis of lattice gauge gravity partition functions, its wave front set analysis.

In quantum field theory, wave front sets give information on the amplitudes with large incoming and outgoing momenta, that is, they describe the high energy part of the theory. In our case the large momentum behavior corresponds to the limit of large boundary geometries.

We showed that wave front sets behave in a geometrically natural way under composition, enabling us to obtain closure and parallel transport equations. This allows us to derive equations for the wave front set of the full ampltiude from that of its weights in a straightforward way. We can then make statements about the entire partition function in the limit of large boundary geometries, without any further approximations. 

We were able to reproduce all the existing results on geometricity, including known issues like ultra-locality in the model of Barrett and Crane.

We could go significantly beyond what was possible before by also giving rigorous, if conditional, geometricity results on the entire partition function on arbitrary manifolds. These methods thus allow for immediate plausibility checks for a wide range of models.

As a first application we clearly demonstrate that the most studied models, all of which include the so called Immirzi parameter $\gamma$ suffer from accidental curvature constraints, strongly disfavoring them. This is a significant refinement of the flatness issue raised by Bonzom in \cite{Bonzom:2009wm,Bonzom:2009hw}. We illuminate its geometric origin and propose a modified partition function that is likely to encode the correct geometricity constraints.

Further development of this method into a full symbolic calculus of distributions on homogeneous spaces, which would be of independent mathematical interest, will allow us finer control of the distributions, establishing whether the subset of geometric configurations on which the partition function is peaked is that satisfying the Regge equations of discrete gravity.

The methods introduced allow us to cast the problem of regularizing the partition function in terms of an extension problem of distributions, closely mirroring the first step of renormalization analysis for quantum field theory on curved space times.

In that context the development of a full symbolic calculus, extending the properties presented here, will dramatically simplify parts of the analysis of quantum fields on homogeneous spaces like de Sitter and anti de Sitter.

In the context of quantum gravity we have provided here equations governing the divergent behavior of the partition function in the conditional of the main result. To follow a renormalization program analogous to that employed successfully in quantum field theory the next step will be to study the ambiguities in the extension of the distribution. At that stage we expect to obtain a clear picture on whether lattice gauge gravity provides a viable approach to the problem of quantum gravity.

\section*{Acknowledgements}

The authors would like to thank Bianca Dittrich, Laurent Freidel, Carlos Guedes, Daniele Oriti, Claudio Perini, Carlo Rovelli, and Lorenzo Sindoni for discussions based on a draft of this letter.

\bibliography{AD-Letter}

\appendix*
\newpage

\begin{widetext}

\section*{Supplemental Information}

\subsection{Wave front sets}

Here we give a fuller account of the structural properties of wave front sets underlying the two properties given in the letter. The set of properties given below either are directly given in \cite{grigis1994microlocal} or follow easily. Together they form a calculus of wave front sets that allows us to study quite arbitrary convolution products of distributions on homogeneous spaces.

For completeness we recall the setup here. We will give general properties of distributions $A \in \calD(M)$ on a smooth compact manifold $M$. The wave front set $\WF(A)$ is a subset of the cotangent bundle on $M$, $\WF(A) \subset T^*M$. We denote the direct product of subspaces of $W^{i} \subset T^*M^{i}$ by $ \times_i W^i \subset \times_i T^*M^i$. We define the sum of subspaces $W_i \subset T^*M$ as $\sum_i W_i = \{(x,p):p=\sum_i p_i, (x,p_i) \in W_i\} \subset T^*M$.


\subsubsection{General properties}

Wave front sets have the following general properties (\cite{lars} chapter 8.1):

\begin{pro}[Multiplication of distributions]\label{pro-mult}
Take $A_i \in \calD(M)$ and denote as $W_{\mbox{cl}} \subset W \subset \times_i T*M$ the subset of $(x^1,x^2,\dots, p^1, p^2, \dots) \in W$ of the form $(x,x,\dots,p^1,p^2,\dots)$ satisfying $\sum_i p^i = 0$.

Then if $$(\times_i \WF(A_i))_{\mbox{cl}} = \times_i \WF(A_i)|_{p^i = 0},$$ then $\prod_i A_i$ exists as a distribution $\prod_i A_i \in {\calD(\times_i M_i)}$ and 
\begin{equation}
 \WF\left(\prod_i A_i\right)\subset \sum_i \WF(A_i)
\end{equation}
\end{pro}

\begin{pro}[Extension]\label{pro-ext}
If $A$ is a distribution in $\calD(M_1 \times M_2)$ defined through a distribution $A' \in \calD (M_1)$ through $A(x_1,x_2) = A'(x_1)$ then $\WF(A) = \WF(A') \times \{0\}_2$. By an abuse of notation we will often denote both distributions by the same letter if it is obvious on which manifold they act.
\end{pro}

\begin{pro}[Integration]\label{pro-int}
Let $A\in D(M_1\times M_2)$ and $M_2$ be compact. Then $A' = \int dx_2 A $ exists as a distribution and $\WF(A') \subset \pi_1 \WF(A)|_{p_2 = 0}$, where $\pi_1$ is the projection on the first component of $T^*M_1 \times T^*M_2$.
\end{pro}

\begin{pro}[The delta function] \label{pro-delta}Let $N\subset M$ be a smooth submanifold. Let $\delta_N$ be a delta
function of $N$ then
\begin{equation}
 \WF(\delta_N)=\{(x,p)\colon x\in N,\forall {v \in TN}, (p,v)=0\}\cup \{0\}
\end{equation}
\end{pro}

\subsubsection{Pseudo-differential operators}

\begin{pro}[Principal symbols]\label{pro-diff}
Let $C$ be a pseudo-differential operator on $M$, and $c$, a function on $T^*M$, its principal
symbol \cite{grigis1994microlocal}. If $CA$ is smooth then
\begin{equation}
 \WF(A)\subset \{c=0\}\cup \{0\}
\end{equation} On the set where the exterior derviative is non vanishing, $\dd c\not=0$ on $\{c=0\}$, $p \neq 0$, $\WF(A)$ is also invariant under the
hamiltonian flow generated by $c$.
\end{pro}

\begin{pro}[Smoothing]
For $C$ and $A$ as above we have
$$\WF(CA) \subset \WF(A)$$
\end{pro}

\subsubsection{Properties on homogeneous spaces}

On homogeneous spaces we obtain further properties. Let $G$ be a (Lie) group acting smoothly on $M$

\begin{pro}[Invariance] \label{pro-inv} If $G \act A = A$ then $\WF(G \act A) = \WF(A)$. If $G$ is Lie and the action is generated by the vector fields $L\in {\mathfrak g}$ then
\begin{equation}
 \WF(A)\subset \{(x,p)\in T^*M\colon \forall_{L\in{\mathfrak g}}\ (p,L)=0\}
\end{equation}
\end{pro}

\begin{pro}[Inverse] \label{pro-inverse} For $A\in \calD(G)$ as above, let $\tilde{A}(g)=A(g^{-1})$, then
\begin{equation}
 \WF(\tilde{A})=\{(g,p)\colon (g^{-1},-gpg^{-1})\in \WF(A)\}
\end{equation}
\end{pro}

From these properties, the composition properties \ref{pro-parallel} and \ref{pro-closure} can be derived relatively straightforwardly.

\subsection{More details on geometricity}

Here we will give more details on the geometric interpretation of the $\Spin(4)$ and $\spin(4)$ elements that occur in equations \eqref{eq:G-VertexEquations}, \eqref{eq:G-twisted-simplicity}, \eqref{eq:G-EdgeEquations} and \eqref{eq:G-FaceEquations}, as well as their symmetries. Before outlining the classification theorems of \cite{Barrett:2009gg,Barrett:2009as,Hellmann:2010nf} that allow us to characterize part of the solution space of these equations geometrically we will first give an inverse statement, and show how, given a continuous simplex-wise flat geometry on a triangulated 4-manifold, we can construct solutions to (\ref{eq:G-VertexEquations}-\ref{eq:G-FaceEquations}).

\subsubsection{The inverse constructions}

The setup is as in the paper with a triangulation and its dual 2-complex $\CC$. We start with a set of oriented, geometric 4-simplices $\sigma^v$ in $\R^4$ defining an orientation and a simplexwise flat, continuous, non-degenerate geometry on the triangulated manifold. These simplices have boundary tetrahedra $\tau^v_e$, with outward normals $N^v_e$. The tetrahedra intersect at the triangles $t^v_{ee'}$. At these triangles we have area outward normals $A^v_{ee'}$ in the plane of the tetrahedron $\tau^v_e$, which satisfy $A^v_{ee'}\cdot t^v_{ee'} = A^v_{ee'}\cdot N^v_e = 0$ and $|A^v_{ee'}| = |t^v_{ee'}|$.

At the middle of the edges $e$ we now introduce a tetrahedron $\tau^e$ with the same geometry as $\tau^v_e$ and $\tau^{v'}_{e}$, normal to some \be N_v^e = -N_{v'}^e\ee chosen such that the orientation it inherits from the standard orientation on $\R^4$ by reduction with the normal $N^e_v$ matches that of the orientation $\tau^{v}_e$ inherits from $N^{v}_e$. This is possible as we required the 4-simplices to define a consistent orientation on the manifold.  We also again have triangles and area normals, that are now, however, indexed by a face, and called $t^e_f$ and $A^e_f$.

We can now define the holonomies of the discrete connection, $G_{ev} \in \SO(4)$ by requiring that

\be G_{ev} \tau^v_e = \tau^e_v.\ee

From this we immediately have that, for $f$ being the face to which $e$, $v$, and $e'$ are adjacent, the outward normals behave well:

\begin{align}
G_{ev} N^v_e & = N^e_v, \nn\\
G_{ev} A^v_{ee'} &= A^e_f.
\end{align}

We also define $G_{ev} = G_{ve}^{-1}$. We can now also introduce the simplicity rotation $G_{ef}$. These are the interior dihedral rotations of the 4-simplices in the frames of the tetrahedra at the edges. Let $e'$ precede $v$ precede $e$ in the fiducial orientation around the face $f$, then we define:

\begin{align}
G_{ef} t^e_f & = t^e_f, \nn\\
- G_{ve} G_{ef} G_{ev} N^v_{e'}&= N^v_e.
\end{align}

\begin{figure*}[htpb]
 \centering
 \includegraphics[scale=.95]{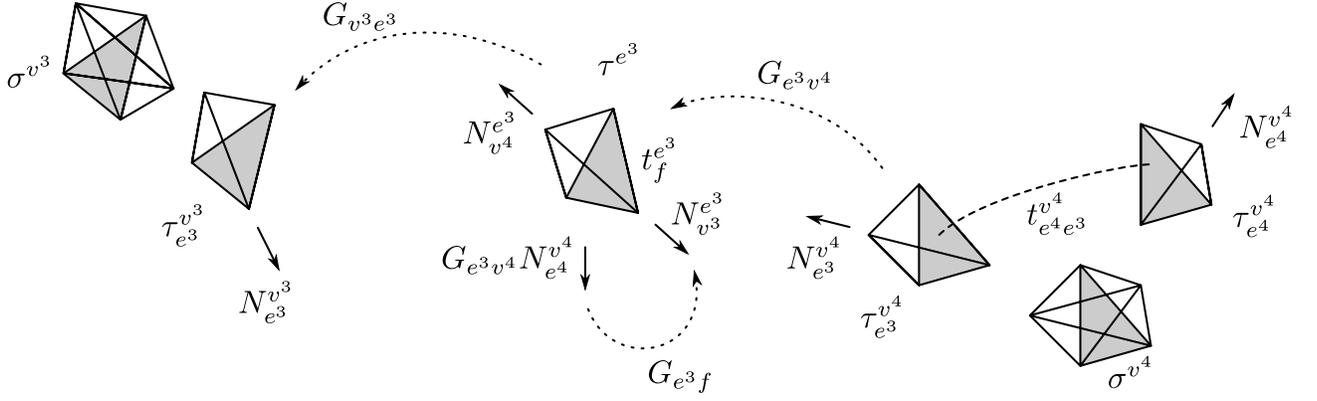}  
 \caption{Geometric quantities at edge $e^3$ in the face $f$, with the sequence of vertices and edges determined by the fiducial orientation being $e^4,v^4,e^3,v^3$.}
 \label{fig-data2}
\end{figure*}

The last line can equivalently be written as $G_{ef} G_{ev} N^v_{e'}= N^e_{v'}$, for $v'$ the vertex succeeding $e$ in the fiducial orientation. It follows that we have $G_{v'e} G_{ef} G_{ev} N^v_{e'}= N^{v'}_{e}$. This arrangement of group elements is given pictorially in figure \ref{fig-data2}.

Consider then the group element around a face with $n$ edges $e \dots e^{(n)}$:

\be G_{f} = G_{ve^{(n)}} G_{e^{(n)}f} G_{e^{(n)}v^{(n-1)}} G_{v^{(n-1)}e^{(n-1)}} G_{e^{(n-1)}f} G_{e^{(n-1)}v^{(n-2)}} \dots G_{v'e} G_{ef} G_{ev}. \ee

This has the property that \be G_f t^{v}_{e(n)e} = t^{v}_{e(n)e}\ee and that \be G_f N^{v}_{e(n)} = N^{v}_{e(n)},\ee thus we have that \be G_f = \id \ee.

From the geometric data we can further construct bivectors, or $\spin(4)$ elements, associated to the triangles. This is easiest using the area normals introduced above:

\begin{align}
p^v_{ee'} &= \hodge N^v_e \wedge A^v_{ee'},\nn\\
p^e_{fv} &= \hodge N^e_v \wedge A^e_f.
\end{align}

Here $\hodge$ is the hodge dual with $\hodge^2 = \id$. By construction these bivectors lie in the plane of the triangles $t^v_{ee'}$ and $t^v_f$ respectively, are oriented,
\begin{align}
p^v_{ee'} &= - p^v_{e'e},\nn\\
p^e_{fv} &= - p^e_{fv'}.
\end{align}
and satisfy the parallel transport equation,
\begin{equation}
G_{ev} p^v_{ee'} = p^e_{fv}.
\end{equation}
It is also immediate to see that they satisfy simplicity,
\begin{equation}
N^e_v \cdot p^e_{fv} = \hodge  N^e_v \wedge N^e_v \wedge A^e_f = 0,
\end{equation}
due to the antisymmetrization in the $\hodge$, and closure
\begin{equation}
\sum_{f \ni e} p^e_{fv} = \hodge N^e_v \wedge \left(\sum_{f \ni e} A^e_f\right) = 0,
\end{equation}
due to the closure of the area outward normals of $\tau^e$.

Further as $G_{ef}$ stabilizes the triangle $t^e_f$ it is generated by the bivector orthogonal to the plane of $t^e_f$, which we can take to be $\xi_{ef} \hodge \hat p^e_{fv}$.

Now we can lift the $\SO(4)$ elements $G_{ev}$ and $G_{ef}$ to $\Spin(4)$ elements $g_{ev}$ and $g_{ef}$ in such a way that we retain $g_f = \id$. THis is the only equation sensitive to the lift. Further we can choose $N^e_v = N^0 = (1,0,0,0)$ for $v$ preceding $e$ in the fiducial orientation. This way we obtain a solution to equations \eqref{eq:G-VertexEquations} till \eqref{eq:G-FaceEquations} from a continuous, simplexwise flat, non-degenerate geometry and orientation on the manifold. Thus we see that geometries occur among the solutions of $\GG_\CC^0$. As we discussed in the letter only a limited set of geometries occur among the solutions of $\GG_\CC^\gamma$ due to the additional accidental curvature constraints.

\subsubsection{The symmetries of the equations}

The equations of $\GG_\CC^\gamma$ allow for a set of symmetries. These can be separated into two types, those that are associated to symmetries of the geometry and those that change the geometric interpretation of the solution. We begin with the former.

\subsubsection{Geometric symmetries}

The first set of symmetries acts simply by rotation at the vertex and the edge. Given a set of elements $g_{v} \in \Spin(4)$ and $g_e \in \SU(2)_{\mbox{diag}}$ we have
\begin{align}
g_{ve} &\rightarrow g_v g_{ve} g_e^{-1},\nn\\
g_{ef} &\rightarrow g_e g_{ef} g_e^{-1},\nn\\
p^v_{ee'} &\rightarrow g_v \act p^v_{ee'},\nn\\
p^e_{fv} &\rightarrow g_e \act p^e_{fv}.
\end{align}

On geometric configurations this corresponds to the transformation
\begin{align}
\sigma^v &\rightarrow g_v \act \sigma^v,\nn\\
\tau^e &\rightarrow g_e \act \tau^e,
\end{align}
which clearly does not change the geometric interpretation.

Further there is a set of symmetries which leaves the geometric content of the configuration untouched but locally changes the orientation. Note that acting with an $\OO(4)$ element $P_v$ on $\sigma^v$ we retain the same geometry. Thus it is still possible to map the tetrahedra $\tau^e$ to the boundary of $P_v \sigma^v$ by an $\SO(4)$ element, however, the normal $N^e_v$ will now not be taken to the outward normal $N^v_e$, but the inward normal $-N^v_e$. Thus the bivectors get transported not to the geometric bivectors of $P_v \sigma^v$, but to the negative bivectors. Calling the reflection with respect to the plane of $\tau^e$, $P_e$, this symmetry thus acts as \begin{align}
G_{ve} &\rightarrow P_v G_{ve} P_e, \nn\\
p^v_{ee'} &\rightarrow P^{-1}_v p^v_{ee'}.
\end{align}
The issue of these orientations is discussed in detail in the literature \cite{Barrett:2009gg,Barrett:2009as,Hellmann:2010nf}.

\subsubsection{Non-geometric symmetries}
The first non-geometric symmetry to consider is the ambiguity in the lifting of the $G_{ev}$ to $\Spin(4)$. This ambiguity can be parametrized by a set of signs $\sigma_{ev}$. If these satisfy $\prod_{e \inf f} \sigma_{ev} = 1$, then they do not change the equations. This can be seen as the different discrete spin structures on $\CC$. Note that the $G_{ef}$ have a canonical lifting given in terms of the $\xi_{ef}$.

The final symmetry relates different $\xi_{ef}$. For convenience we will work with the rescaled $\xi'_{ef} = \frac{\xi_{ef}}{\sqrt{1+\gamma^2}}$. Call $n_\gamma$ the smallest denominator of $\gamma$. Given a solution $\xi'_{ef}$ we immediately obtain another solution as $\xi'_{ef} \rightarrow \xi'_{ef} + m_{ef} n_{\gamma} 4 \pi$ for any $m_{ef} \in \Z$. In other words, $\xi'_{ef}$ should be viewed geometrically as living on a circle of circumferences $n_{\gamma} 4 \pi$.

Furthermore, as only the sum $\sum_{e \in f} \xi'_{ef}$ enters into the equations, and only up to a factor of $4 \pi$, we have a symmetry parametrized by $\tilde{\xi}_{ef}$ satisfying $\sum_{e \in f} \tilde{\xi}_{ef} = 0 \mod 4 \pi$, that acts as
\begin{align}\label{eq:sym-U1}
 \xi_{ef} &\rightarrow \xi_{ef} + \tilde{\xi}_{ef},\nn\\
 g_{ef} &\rightarrow g_{ef} \exp(\tilde{\xi}_{ef} \hat{p}^e_{fv}).
\end{align}

\subsubsection{Geometric reconstruction}

The geometric reconstruction theorems of \cite{Barrett:2009gg,Barrett:2009as,Hellmann:2010nf} classify the solutions of \eqref{eq:G-VertexEquations} and \eqref{eq:G-EdgeEquations} at each vertex. It is shown that these fall into three categories:
\begin{itemize}
 \item Fully degenerate configurations where the edge geometry becomes to two or less dimensional.
 \item $\SU(2) BF$ solutions, which do not define a 4-dimensional geometry but a 4-dimensional $\SU(2) BF$ configuration.
 \item Configurations given by the geometric inverse construction of the preceding subsection, and their image under the orientation reversing symmetry.
\end{itemize}

Given a solution in the last category we can further study the $\xi'_{ef}$ or $g_{ef}$ respectively. Equation \eqref{eq:G-FaceEquations}, fixes the sum of $\xi'_{ef}$ to be $\Theta_f$ mod $4 \pi$. We can then use the symmetry \eqref{eq:sym-U1} to fix the $\xi'_{ef}$ to be the interior dihedral angles of the preceding 4-simplex, thus completing the reconstruction.

\end{widetext}

\end{document}